\documentclass[english,aps,twocolumn,prb,amssymb,amsmath,superscriptaddress]{revtex4-1}
\usepackage{graphicx,color}
\usepackage[bookmarks]{hyperref}
\usepackage[T1]{fontenc}

\newcommand{\be}{\begin{equation}}
\newcommand{\ee}{\end{equation}}

\begin{document}

\title{Dynamical quantum phase transitions in collapse and revival oscillations of a quenched superfluid}
\author{Mateusz Lacki}
\affiliation{Instytut Fizyki imienia Mariana Smoluchowskiego, Uniwersytet Jagiellonski, Lojasiewicza 11, 30-048 Krakow, Poland}
\author{Markus Heyl}
\affiliation{Max-Planck-Institut f\"ur Physik komplexer Systeme, 01187 Dresden, Germany}

\begin{abstract}
In this work we revisit collapse and revival oscillations in superfluids suddenly quenched by strong local interactions for the case of a one-dimensional Bose-Hubbard model.
As the main result we identify the inherent nonequilibrium quantum many-body character of these oscillations by revealing that they are controlled by a sequence of underlying dynamical quantum phase transitions in the real-time evolution after the quench, which manifest as temporal nonanalyticities in return probabilities or Loschmidt echos.
Specifically, we find that the time scale of the collapse and revival oscillations is, firstly, set by the frequency at which dynamical quantum phase transitions appear, and is, secondly, of emergent nonequilibrium nature, since it is not only determined by the final Hamiltonian but also depends on the initial condition.
\end{abstract}

\date{\today}

\maketitle


\emph{Introduction.---} 
Starting from the observation of collapse and revival  oscillations for a Bose-Einstein Condensate matter wave~\cite{Greiner2002} the field of nonequilibrium quantum many-body physics has seen a rapid development. Experiments in quantum simulators, such as ultra-cold atoms or trapped ions among others~\cite{Bloch2008mb,Bloch2012qs,Blatt2012,Geogescu2014}, have in the meantime observed various inherently dynamical quantum phenomena such as prethermalization~\cite{Gring2012,Neyenhuis2017,2018Singh}, particle-antiparticle production in lattice gauge theories~\cite{Martinez2016}, dynamical quantum phase transitions~\cite{Jurcevic2016,zhang2017observation,Flaschner2018},  many-body localization~\cite{Schreiber2015,Smith2016,choi2016exploring}, or discrete time crystals~\cite{Zhang2017,Choi2017}.
Within the anticipated collapse and revival experiment a Bose-Einstein condensate (BEC) is suddenly quenched by strong local on-site interactions, which leads to a periodic decay and reappearance of a peaked structure in the bosonic momentum distribution characteristic of a BEC \cite{Greiner2002} or the visibility of interference patterns~\cite{2006JPhB...39S.199A,2007PhRvL..98t0405S,Bloch2010}.
While many aspects of this experiment have been theoretically addressed~\cite{Kollath2007,Fischer2008,Johnson2009,Wolf2010,Schachenmayer2011,Fischer2011}, it has remained elusive whether there is a general dynamical principle underlying these collapse and revival oscillations, which can explain some central questions that are still unanswered such as concerning the time scale of these oscillations.

\begin{figure}
	\includegraphics{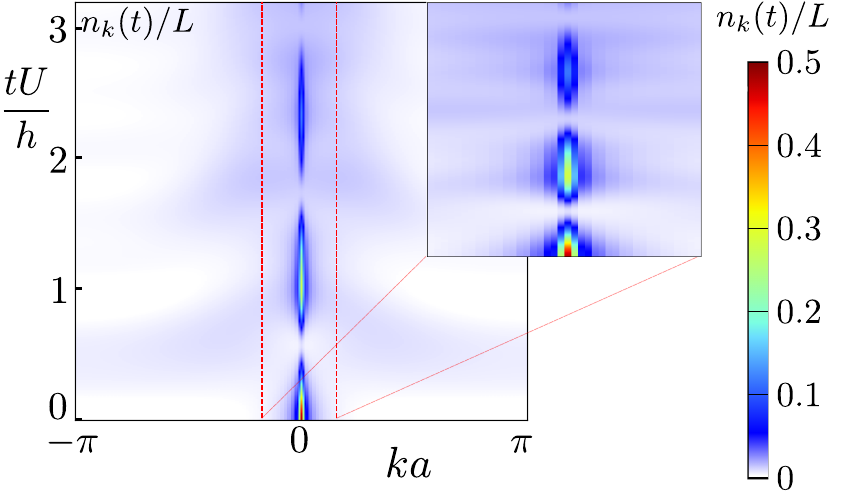}
	\caption{\label{fig:1} Collapse and revival oscillations in the real-time evolution of the quasimomentum distribution $n_k(t)$ for a superfluid suddenly quenched by strong local interactions. The system is initially prepared in the superfluid ground state of a one-dimensional Bose-Hubbard model with $L=120$ lattice sites for modest initial interactions $U_i$ with $s_i = J_i/U_i = 0.36$ where $J_i $ denotes the initial hopping amplitude. After quenching the system suddenly at time $t=0$ to $s_f = J_f/U = 0.05$, the system exhibits a periodic sequence of collapse and revival oscillations of the zero quasimomentum peak. The inset shows a magnification of the area at low quasimomenta enclosed by the red rectangle.}
\end{figure}

In this work we reexamine the collapse and revival oscillations for the case of superfluid order in a one-dimensional Bose-Hubbard model (BHM) subject to a sudden quench of strong local interactions.
As for the collapse and revival experiment we observe a periodic sequence of decay and reappearance of the zero-momentum peak in the bosonic distribution function, see Fig.~\ref{fig:1}.
It is the main result of this work to identify the collapse and revival oscillations as a genuine nonequilibrium quantum many-body problem phenomenon (i) by relating them to dynamical quantum phase transitions~\cite{heyl_dynamical_2013,2016LTP....42..971Z,heyl2018dynamical,2018arXiv181102575H}, and (ii) by revealing the origin and emergent nonequilibrium nature of the associated time scale.
Specifically, we find that the collapse and revival oscillations are controlled by a sequence of underlying DQPTs, that are characterized by a nonanalytic behavior as a function of time contained in the Loschmidt amplitude
\be
	\mathcal{G}(t) = \langle \psi_0 | e^{-iHt} | \psi_0 
	\rangle \, ,
	\label{eqn:LochschmidtEcho}
\ee
where $|\psi_0\rangle$ denotes the initial condition and $H$ the Hamiltonian driving the quantum real-time dynamics.
$\mathcal{G}(t)$ quantifies the amplitude to return to the initial condition and is therefore a natural measure for the collapse and revival of the properties of the initial state such as the zero-momentum peak.
As the main result of this work we observe that the time scale $t_\ast$ for the collapse and revival oscillations matches the periodicity at which the system experiences DQPTs.
In this way we provide an explanation of the time scale for these oscillations and link them to a phenomenon that provides general principles of quantum real-time evolution.
Importantly, we show that $t^\ast$ is an emergent nonequilibrium time scale without an equilibrium analog depending both on the initial condition and the final Hamiltonian parameters and therefore $t^\ast$ is not set, for example, just by the gap of the final Hamiltonian.


\emph{Model and setup.---} 
The BHM describes interacting bosons on a lattice close to the aforementioned experiment \cite{Greiner2002}. We take the underlying lattice to be one-dimensional yielding the following Hamiltonian:
\be
H_{\textrm{BHM}}(J,U)=-J\sum_{i=1}^{L-1} a_i^\dagger a_{i+1}+a_{i+1}^\dagger a_{i}+\frac{U}{2}\sum_{i=1}^{L} n_i(n_i-1).
\label{eqn:BHM}
\ee
where $a_i$ is the annihilation operator for a boson on site $i$ and $n_i=a^\dagger a_i$ the corresponding occupation. The lattice consists of $L$ sites and for convenience we choose open boundary conditions, which, however, has no influence on our main results.
The properties of the Hamiltonian $H_{\textrm{BHM}}(J,U)$ are determined by the dimensionless ratio $s=J/U$ between hopping amplitude $J$ and interaction strength $U$.
At zero temperature, a quantum phase transition of Kosterlitz-Thouless type occurs for unit filling at $s_c \approx 0.297$ separating a Mott insulating (MI) for $s<s_c$ from a  superfluid (SF) phase for $s>s_c$~\cite{Kuhner1998,Zakrzewski2008,Rachel2012,Carrasquilla2013,Krutitsky2016}.

To model the collapse and revival oscillations we initialize the system in a superfluid ground state $|\psi_0\rangle$ of the BHM for $s>s_c$.
At time $t=0$ the interaction is suddenly quenched to a large value with $s<s_c$ inducing nonequilibrium real-time dynamics which can be formally solved by
\be
	|\psi_0(t) \rangle = e^{-iHt} |\psi_0\rangle \, ,
\ee
where $H$ denotes the final quenched Hamiltonian.
Just as in the collapse and revival experiment \cite{Greiner2002} and in related theory works~\cite{Zoller2011} we study the properties of this quenched system via the quasimomentum distribution
\be
n_{k}(t)=\langle \psi_0(t) | n_k| \psi_0(t) \rangle \, ,
\ee
where $n_k=a_k^\dagger a_k$ with $a_k=1/\sqrt{L} \sum_j e^{ijk} a_j$ for $k=-\pi,-\pi+2\pi/L,\ldots ,\pi$.
The dynamics of $n_k(t)$ is shown for a representative set of parameters in Fig~\ref{fig:1}.
Since the system is initially prepared in a superfluid state, the quasimomentum distribution shows a macroscopic occupation at zero momentum $k=0$.
Upon quenching strong local interactions $U$, $n_k(t)$ exhibits on transient time scales a decay of the superfluid signature at $k=0$ leading to a spread of occupation across the whole Brillouin zone, see the blue shade in Fig.~\ref{fig:1}. 
After this decay, the zero-momentum peak, however, reappears again.
This sequence of collapses and revivals continues periodically.
In the limit of very strong interactions $s\to 0$ the time-dependence distribution $n_k(t)$ is perfectly periodic in time.
For nonzero $0<s\ll s_c$ instead, the peak height at $n_{k=0}(t)$ exhibits an additional decaying envelope for the subsequent revivals.
Still, if the decay time is longer than revival time, the periodic character of the collapse and revival oscillations is clearly present. This point will be discussed in greater detail later in the main text.

We have obtained this data and the results in the remainder of this work using numerical simulations based on Matrix Product State (MPS) techniques.
The ground states of the BHM model show sufficiently low entanglement entropy to be accurately represented by MPS~\cite{Schollwock2011} with a small bond dimension~\footnote{we consider bond dimension of $m=160$ and $n_{\textrm{\footnotesize max}}=7$ bosons per site }.
The initial ground state we compute by means of the Density Matrix Renormalization Group (DMRG) algorithm \cite{DMRG}. 
In order to perform the real-time evolution of the initial state under the quantum quench we use the Time-Evolving Block Decimation (TEBD) \cite{Vidal2003}.
A temporal linear growth of the entanglement entropy limits the maximally achievable evolution time $T$ to $T=20/U$, up to which our numerics remains accurate.
For TEBD we have used a 4th order Trotter formula for factorization of the time evolution operator with a time step $\delta t < 0.002/U$. We also note that an MPS representation of vectors allows for accurate computation of very small overlaps, necessary to compute $\mathcal{G}(t)$ without resorting to high precision linear algebra \cite{Schollwock2011,DMRG}. This is in contrast to the direct evaluation in Fock basis, where subtraction of O(1) term introduces a numerical problem due to finite resolution of double precision floating point numbers. 


\emph{Dynamical quantum phase transitions.---}
As outlined in the introduction it is the purpose of this work to link the collapse and revival oscillations as seen in Fig.~\ref{fig:1} to DQPTs and therefore to a genuine nonequilibrium critical phenomenon.
Before addressing this connection in detail, let us first outline some basic properties of DQPTs.
The theory of DQPTs provides an extension for the concept of phase transitions to the nonequilibrium dynamical regime~\cite{heyl_dynamical_2013,heyl2018dynamical}.
While equilibrium transitions are driven by external control parameters such as temperature or pressure, DQPTs are caused solely by the system's internal unitary dynamics.
The central object is the Loschmidt amplitude $\mathcal{G}(t)$, see Eq.~(\ref{eqn:LochschmidtEcho}), and the related probability $\mathcal{L}(t) = |\mathcal{G}(t)|^2$ called the Loschmidt echo.
Formally, $\mathcal{G}(t)$ resembles equilibrium partition functions at complex parameters~\cite{heyl_dynamical_2013,heyl2018dynamical}.
Accordingly, it is suitable to introduce dynamical analogs to free energy densities.
In the following we will consider mainly the dynamical free energy density $\lambda(t)$ corresponding to the Loschmidt echo $\mathcal{L}(t)$ defined as:
\be
	\lambda(t) = -\frac{1}{N} \log \big[ |\mathcal{G}(t)|^2\big] \, .
\ee
As conventional free energy densities can become nonanalytic at phase transitions, so can the dynamical counterpart $\lambda(t)$ in the thermodynamic limit, but now at critical times which is the defining feature of DQPTs.
Recently, DQPTs and their signatures have been observed experimentally in quantum simulators realized in trapped ions~\cite{Jurcevic2016}, ultra-cold atoms~\cite{Flaschner2018,2018arXiv180209229H}, quantum walks~\cite{Wang2018,Xu2018}, nanomechanical oscillators~\cite{Tian2018}, and superconducting qubits~\cite{Guo2018}.
It has been shown that many important properties of equilibrium transitions beyond mere nonanalytic behavior are also shared by DQPTs.
This includes, for example, their robustness against symmetry-preserving perturbations~\cite{Karrasch2013,Kriel2014,Heyl2015dq,Sharma2015}.
Moreover, dynamical order parameters have been constructed~\cite{Budich2015,Sharma2016,Bhattacharya2017,Bhattacharya2017b,Bhattacharya2017c,Flaschner2018,HeylBudich2017} and measured~\cite{Flaschner2018,Wang2018,Xu2018,Guo2018}, Landau theories have been formulated~\cite{Trapin2018}, as well as scaling and universality have been identified~\cite{Heyl2015dq,Trapin2018} for specific models. 
%


\begin{figure}
	\includegraphics[width=86mm]{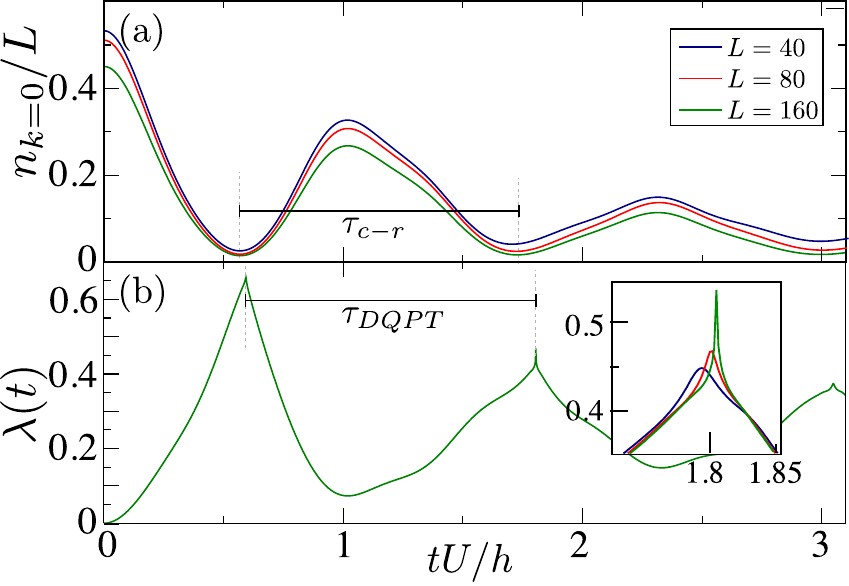}
	\caption{(a) Real-time evolution of the zero-quasimomentum peak $n_{k=0}(t)$ of the bosonic distribution function for different system sizes and the same parameter values as in Fig.~\ref{fig:1} displaying the collapse and revival oscillations. The dashed lines indicate the position of the first two minima of $n_{k=0}(t)$ which defines the time scale $\tau_{CR}$. (b) Dynamics of the Loschmidt echo rate function $\lambda (t)$. The dashed lines mark the location of the first two DQPTs and their temporal distance gives the time scale $\tau_{\textrm{\footnotesize DQPT}}$ of the appearance of DQPTs.The inset shows the sharpening for increasing system size $L$. }
	\label{fig:DQPTmomentum}
\end{figure}

\emph{Results.---}
Along the lines of our main goal of connecting collapse and revival oscillations to DQPTs, we compare in Fig.~\ref{fig:DQPTmomentum} the dynamics of the zero-quasimomentum occupation $n_{k=0}(t)$ to the evolution of the dynamical counterpart of the free energy density $\lambda(t)$ for the same parameters as in Fig.~\ref{fig:1}.
While $n_{k=0}(t)$ shows clearly the discussed collapse and revival oscillations of the superfluid order, $\lambda(t)$ exhibits a sequence of sharp structures in its real-time evolution. 
Moreover, the location of these sharp features in time appears correlated with the minima in $n_{k=0}(t)$, as we will discuss more quantitatively below.
The sharp structures in time appearing in $\lambda(t)$ become sharper for increasing system size $L$, as we show for one case in the inset of Fig.~\ref{fig:DQPTmomentum}(b).
As a consequence, we conclude that these features eventually turn into nonanalytic kinks in the thermodynamic limit, which is the defining property of a DQPT.
Importantly, the system experiences not only a single DQPT in consequence of the considered quantum quench, but rather a whole sequence, the first three of which are contained in the time interval shown in Fig.~\ref{fig:DQPTmomentum}.
Clearly, finite-size effects for the kinks become stronger at larger times, so that we restrict our analysis to involve only the first two DQPTs.
Further, finite-size effects also appear to become more important upon increasing $s_i$, i.e., when choosing the initial superfluid at weaker interactions, such that we limit ourselves to $s_i\leq 1$ in the remainder.
Comparing the time traces of $n_{k=0}(t)$ to $\lambda(t)$ in Fig.~\ref{fig:DQPTmomentum} already suggests a correlation between the time of collapse, identified with a local minimum of $n_{k=0}(t)$, and the occurrence of a DQPT.
Guided by this observation we will now study the connection between collapse and revival oscillations with DQPTs more quantitatively by comparing directly the time scales: $\tau_{CR}$ for the periodic decay of the zero-quasimomentum peak  and $\tau_{\textrm{\footnotesize DQPT}}$ for the sequence of DQPTs.
We extract $\tau_{CR}$ and $\tau_{\textrm{\footnotesize DQPT}}$ as indicated in Fig.~\ref{fig:DQPTmomentum} via the temporal difference between two minima in $n_{k=0}(t)$ and between two DQPTs, respectively. 

\begin{figure}
	\includegraphics{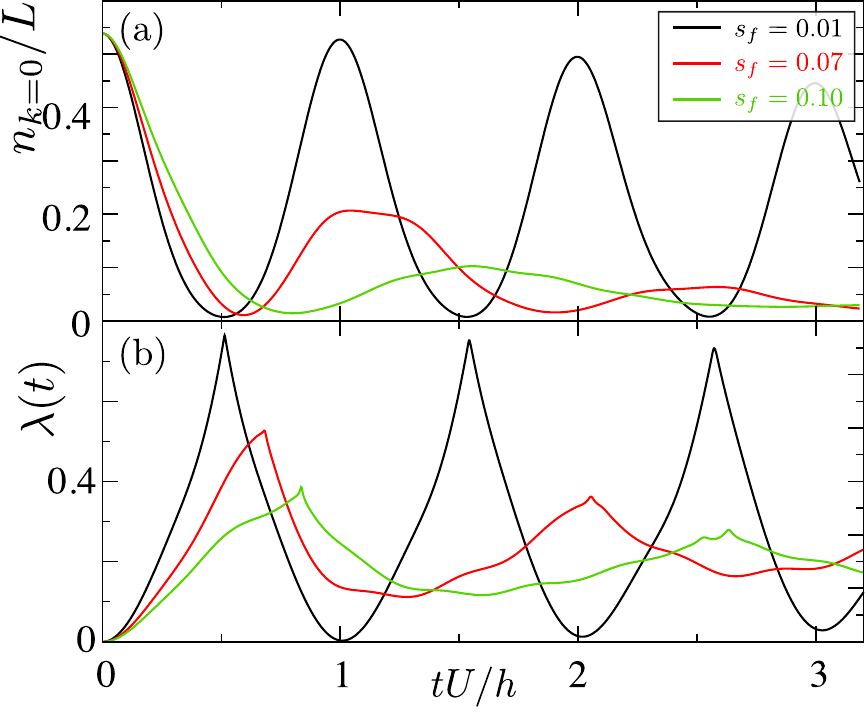}
	\caption{Dependence of the collapse and revival oscillations as well as the DQPTs as a function of the final Hamiltonian parameter $s_f = J_f/U_f$ at $L=120$ and $s_i=0.36$. (a) The dynamics of the zero-quasimomentum peak $n_{k=0}(t)$ exhibits stronger decay and longer oscillation time scales upon increasing $s_f$. (b) Evolution of the Loschmidt echo rate function $\lambda(t)$ shows also a shift towards larger times of the DQPTs as well as an increased damping for larger $s_f$.}
	\label{fig:parameterdependence}
\end{figure}

\begin{figure}
	\includegraphics{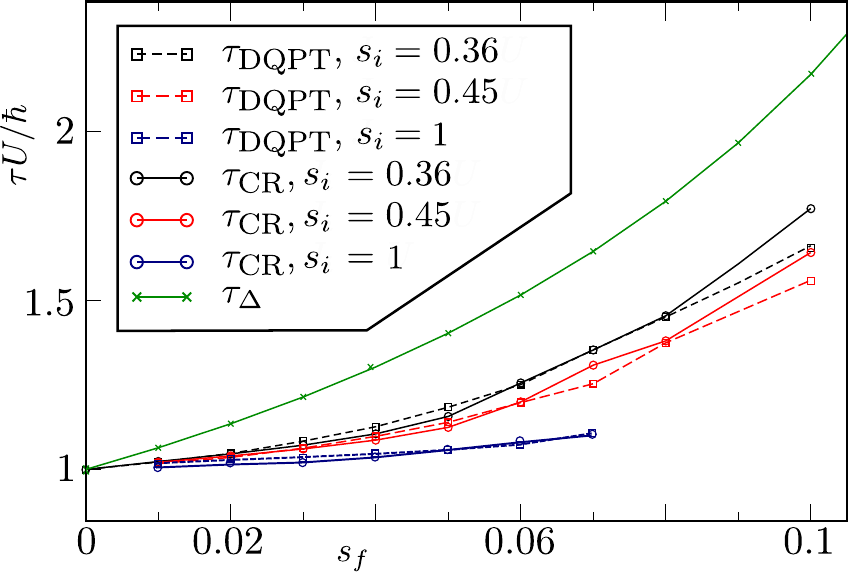}
	\caption{Comparison of the timescale for the collapse and revival oscillations $\tau_{{\footnotesize CR}}$ and the timescale $\tau_{\textrm{\footnotesize DQPT}}$ for DQPTs for different initial conditions $s_i$ as a function of the final Hamiltonian parameter $s_f$ at $L=120$. The dashed and solid lines show $\tau_{{\footnotesize CR}}$ and $\tau_{\textrm{\footnotesize DQPT}}$, respectively. The different colors blue, red, and black refer to different initial conditions $s_i=0.36,0.45,1.0$. The fine-dashed curve displays the equilibrium timescale $\tau_{\Delta} = 2\pi/\Delta$ set by the gap $\Delta$ of the final Hamiltonian.}
	\label{fig:timescales}
\end{figure}

In Fig.~\ref{fig:parameterdependence}(a) we show our numerically obtained data for the zero-quasimomentum occupation $n_{k=0}(t)$ upon varying the parameter $s_f$ of the final Hamiltonian for a fixed initial condition of $s_i=0.36$
For very small $s_f$ the collapse and revival oscillations are very prominent.
Upon increasing $s_f$ the oscillation period grows while at the same moment also the decay of the signal increases.
For $s_f=0.1$ this reduces the visibility of the collapse and revival pattern already rather significantly such that it becomes difficult to resolve $\tau_{CR}$ for even larger $s_f$.
This becomes even more pronounced for larger initial $s_i$ so that we limit our discussion in the following only to those cases where the oscillations are significantly visible for the first two collapses.
In Fig.~\ref{fig:parameterdependence}(b) we show additionally our obtained data for the dynamical free energy density $\lambda(t)$.
In this case the ratio of maximal to minimal value of the $\lambda(t)$ decreases as well with larger $s_f$. This alone does not impair our ability to extract $\tau_{\textrm{\footnotesize DQPT}}$ even for relatively large $s_f\approx 0.1$ unlike $\tau_{\textrm{\footnotesize CR}}$. 

What becomes a challenge is that typically for larger $s_f$ the function $\lambda(t)$ has a tendency to become smoother, requiring larger values of $L$ to locate a non-analytic peak.
 The time at which it occurs is also defined by the fact that except narrow time intervals around DQPTs, the function $\lambda(t)$ is practically $L$-independent. 
 The size of these intervals shrinks as $L$ is increased (see inset of Fig. \ref{fig:DQPTmomentum}(b)). This provides an estimate of location of the DQPT.


In Fig.~\ref{fig:timescales} we now compare the time scales $\tau_{CR}$ and $\tau_{\textrm{\footnotesize DQPT}}$ as defined in Fig.~\ref{fig:DQPTmomentum}.
We plot these as a function of the final parameters $s_f$ for three initial conditions given by $s_i=0.36,0.45,1$.
In addition we include as a reference also the time scale $\tau_\Delta=2\pi \hbar /\Delta$ associated with the gap $\Delta$ of the final Hamiltonian.
For $s_f=0$ the collapse and revival oscillations occur with a period $\tau_{CR}=2\pi \hbar/U$~\cite{Greiner2002,Zoller2011}, which is just a consequence of the discrete equidistant spectrum of the Hamiltonian $H_f=U\sum_{l} n_l(n_l-1)$.
Consequently, all the time scales have to agree in this limit, as one can also clearly identify in Fig.~\ref{fig:timescales}.
Upon increasing $s_f$ one can see corrections to the $s_f=0$ limit.
Here, $\tau_{CR}$ and $\tau_{\textrm{\footnotesize DQPT}}$ stay close to each other while $\tau_\Delta$ deviates significantly.
However, we observe slight deviations between $\tau_{CR}$ and $\tau_{\textrm{\footnotesize DQPT}}$ which we attribute to two possible origins:
(i) finite-size effects that lead to slight shifts of both of the time scales, as one can already see from Fig.~\ref{fig:DQPTmomentum}.
(ii) the reduced visibility of the collapse and revival signal upon increasing $s_f$ as discussed before.
Clearly, however, $\tau_{CR}$ and $\tau_{\textrm{\footnotesize DQPT}}$ go hand in hand with each other whereas $\tau_\Delta$ behaves completely differently.
Further, $\tau_{CR}$ and $\tau_{\textrm{\footnotesize DQPT}}$ exhibit a marked influence of the initial conditions.
All these observations suggest that the collapse and revival oscillations are associated with an \emph{emergent nonequilibrium time scale}, which does not exhibit an equilibrium counterpart.
Since this nontrivial time scale also appears in the nonanalyticities of the Loschmidt echo, we conclude that these oscillations are controlled by a sequence of underlying DQPTs.
This interpretation aligns well with previous works that have found a relation between order parameter dynamics and DQPTs in other models~\cite{heyl_dynamical_2013,2014PhRvL.113t5701H,Budich2015,2017PhRvB..96m4313W,2018PhRvL.120m0601Z,2018arXiv180807874H} as well as in a recent experiment~\cite{Jurcevic2016}.

\emph{Concluding discussion.---}
Closed nonequilibrium quantum many-body systems cannot be characterized by means of thermodynamics.
On the one hand this allows to relax equilibrium constraints such as the equal a priori probability of the microcanonical ensemble, which can lead to quantum states with novel properties, time crystals for instance~\cite{2016PhRvL.116y0401K,2016PhRvL.117i0402E,sacha2015modeling,Zhang2017,Choi2017,Sacha2017}.
On the other hand this makes the theoretical description challenging, since now it is generally not sufficient to characterize properties on the level of Hamiltonians but rather on the level of time evolution operators $U(t)$.
This can be seen for instance from one central result of our work, which is that the collapse and revival oscillations are described by an emergent nonequilibrium time scale, that is not only determined by the properties of the final Hamiltonian but also depends on the initial condition.
Most importantly, studying $U(t)$ adds an additional scale into the problem, which is time $t$ itself.
The theory of DQPTs provides a general framework to incorporate time explicitly and to study the properties of time evolution operators $U(t)$, since for instance the central object $\mathcal{G}(t)$ can be interpreted as a matrix element of $U(t)$.
Quenches and DQPTs in other one-dimensional models exhibiting superfluid to Mott insulator transitions in equilibrium have been studied recently~\cite{2017NJPh...19k3018F}, where no apparent connection between the zero-quasimomentum distribution and DQPTs has been found.
Compared to the present work and others where a close connection between the order parameter dynamics and DQPTs has been observed~\cite{heyl_dynamical_2013,2014PhRvL.113t5701H,Budich2015,2017PhRvB..96m4313W,2018PhRvL.120m0601Z,2018arXiv180807874H,Jurcevic2016}, the order parameters of the models in Ref.~\cite{2017NJPh...19k3018F} exhibit a particularly simple structure due to the underlying integrability of the considered systems, suggesting that our results might not generalize to such simple models of superfluid to Mott insulator transitions.
The initial experiment has been performed in a three-dimensional optical lattices~\cite{Greiner2002}.
Our theoretical considerations use a chain instead, leading immediately to the question of how our results might extend to higher dimensions and, in particular, whether the oscillations are similarly related to DQPTs.
This cannot be addressed by means the MPS formalism used here, but might be within reach of projected-entangled pair states (PEPS)~\cite{2008AdPhy..57..143V} in the future.
However, let us emphasize that the Loschmidt echo still naturally provides a measure for the departure from and return to the initial superfluid state and therefore of the collapse and revival oscillations.

While collapse and revival oscillations can be experimentally accessed straightforwardly using time-of-flight imaging~\cite{Bloch2008mb}, measuring Loschmidt amplitude or echos is a challenge.
Recently, however, the return probability, i.e., Loschmidt echo, for condensed bosons in an ultra-cold atom setup has been estimated~\cite{2018Singh}, which gives hope that our theoretical predictions might become observable in the near future.
%


\emph{Acknowledgments.---} We thank Wilhelm Zwerger for important initializing discussions that led to this work as well as Dominique Delande and Jakub Zakrzewski for the TEBD code.
This work was realized under National Science Center (Poland) project 2016/23/D/ST2/00721 and was supported in part by PL-Grid Infrastructure. Financial support by the Deutsche Forschungsgemeinschaft via the Gottfried Wilhelm Leibniz Prize program is gratefully acknowledged.


\bibliographystyle{apsrev4-1}
%

\end{document}